\documentclass[
prd
,showpacs,amssymb,superscriptaddress,aps
]{revtex4}
\usepackage{graphicx}
\usepackage{color}
\input{epsf}

\usepackage{amsmath,amssymb}
\usepackage{bm}
\usepackage{times}
\usepackage{ulem}

\newcommand{\dalm}{\kern1pt\vbox{\hrule height 0.9pt\hbox{\vrule width 0.9pt
\hskip 2.5pt\vbox{\vskip 5.5pt}\hskip 3pt\vrule width 0.3pt}\hrule height 0.3pt}
\kern1pt}

\newcommand{\gsim}{\, \raisebox{-0.8ex}{$\stackrel{\textstyle >}{\sim}$ }}
\newcommand{\lsim}{\, \, \raisebox{-0.8ex}{$\stackrel{\textstyle <}{\sim}$ }}

\begin{document}



\title{Universal relations between the quasinormal modes of neutron star and tidal deformability}

\author{Hajime Sotani}
\email{sotani@yukawa.kyoto-u.ac.jp}
\affiliation{Astrophysical Big Bang Laboratory, RIKEN, Saitama 351-0198, Japan}
\affiliation{Interdisciplinary Theoretical \& Mathematical Science Program (iTHEMS), RIKEN, Saitama 351-0198, Japan}

\author{Bharat Kumar}
\affiliation{Department Of Physics \& Astronomy, NIT, Rourkela-769008, India}

\date{\today}

\begin{abstract}
Universal relations that are independent of the equation of state (EOS) for neutron star matter are valuable, if they exist, for extracting the neutron star properties, which generally depend on the EOS. In this study, we newly derive the universal relations predicting the gravitational wave frequencies for the fundamental ($f$), the 1st pressure ($p_1$), and the 1st spacetime ($w_1$) modes and the damping rate for the $f$- and $w_1$-modes as a function of the dimensionless tidal deformability. In particular, with the universal relations for the $f$-modes one can predict the frequencies and damping rate with less than $1\%$ accuracy for canonical neutron stars. 
\end{abstract}

%
\maketitle


\section{Introduction}
\label{sec:I}
Thanks to GW170817 event, the first detection of the gravitational waves from binary neutron star merger has been initiated to explain the nature of dense matter at supranuclear densities inside the neutron stars as well as placed the bound on the canonical neutron star tidal deformability \cite{gw170817}. 
Subsequently, the 2nd detection of gravitational waves from the binary neutron star merger, GW190425, has succeeded \cite{GW190425}. The event rate from the binary neutron star merger will become more and more in the future observations, owing to the improvement of the gravitational wave detectors' sensitivity. 
Also, GW170817 measurement has constrained the equation of state (EOS) at twice nuclear saturation density \cite{gw170817a}. However, due to the nuclear saturation properties, it is quite difficult for obtaining the nuclear information for a high density region via terrestrial experiments, while the density inside the neutron star significantly exceeds the nuclear saturation density \cite{ST83}. This is a reason why the EOS for neutron star matter is not fixed yet, although various EOSs have been theoretically proposed up to now. One of the possibility for understanding the EOS for dense matter must be astronomical observations. In practice, owing to the discovery of the $2M_\odot$ neutron stars, some of the soft EOSs, with which the predicted maximum mass does not reach the observed mass, have been ruled out \cite{D10,A13,C20}. The GW170817 event also tells us the constraint on the $1.4M_\odot$ neutron star radius \cite{Annala18}, from which too still EOSs may be unsuitable. In addition, the relativistic effect, i.e., a light bending due to the strong gravitational field involved by a neutron star, enables us primarily to probe the neutron star compactness, $M/R$, with mass $M$ and radius $R$, which in turn restricts the EOS (e.g., \cite{PFC83,LL95,PG03,PO14,SM18,Sotani20a}). In fact, the Neutron star Interior Composition Explorer (NICER), which is an x-ray timing and spectroscopy instrument on the International Space Station, is now operating and has already given us the constraint on the mass and radius of the millisecond pulsars, i.e., PSR J0030+0451 \cite{Riley19,Miller19} and PSR J0740+6620 \cite{Riley21,Miller21}. Such observables can be obtained from the neutron star EOS (for example, EOS based on the relativistic framework and Skyrme type interaction) by solving equations for hydrostatic equilibrium, the so-called Tolman-Oppenheimer-Volkoff equation. Further, comparing astrophysical data puts important constraints on the EOS models and, hence, the nature of the dense matter of neutron stars.

 As another possibility for getting the neutron star properties, the asteroseismology is also powerful method, which is a similar technique to seismology on the Earth and helioseismology on the Sun. That is, since the oscillation frequency of the object strongly depends on its interior information, one can inversely extract the object's properties by observing the corresponding frequency. For example, by identifying the magnetar quasi-periodic oscillations with the crustal oscillations in  neutron stars, the crust EOS has been constrained \cite{GNHL2011,SNIO2012,SIO2016}. By observing the gravitational waves from the neutron star, one may see the neutron star mass, radius, and EOS (e.g., \cite{AK1996,AK1998,Benhar99,STM2001,SH2003,Benhar04,TL2005,Lau10,SYMT2011,PA2012,BGN13,DGKK2013,CdK15,Sotani20b,Sotani21}). Recently, there are attempts for understanding the gravitational wave signals appearing in the numerical simulation for core-collapse supernovae with this technique (e.g., \cite{FMP2003,FKAO2015,ST2016,SKTK2017,MRBV2018,SKTK2019,TCPOF19,SS2019,ST2020,STT2021}).

In any case, the universal relation independently of the EOS is very important, if it exits. This is because it is generally very difficult for extracting the information of neutron star from the direct observation(s)  of gravitational wave frequency, due to  the uncertainty in the EOS for neutron star matter. Up to now, a few universal relations for eigenfrequencies from the neutron stars have been derived, i.e., the fundamental ($f$) mode frequency as a function of the neutron star average density, the $f$-mode frequency as a function of the neutron star compactness, and the spacetime ($w$) mode frequency and damping rate as a function of the stellar compactness \cite{AK1996,AK1998,Benhar99,Benhar04,TL2005,Lau10,BGN13,CdK15,Sotani21}. If the characteristics of the gravitational waves are observationally determined, these universal relations will be helpful for constraining the global properties of the neutron stars and then be used to extract the information of the underlying EOS of dense neutron-rich nuclear matter. In this study, we newly find the other universal relations as a function of the dimensionless tidal deformability, $\Lambda$, focusing on the feature that the relation between the neutron star compactness and $\Lambda$ is almost independent of the EOS. With respect to the universal relation between the quasinormal modes of neutron star and tidal deformability, several studies have already been done with the $f$-modes \cite{Chan14,Wen19,LBN21} and with the $w_1$-modes \cite{MG19,Benitex21}. In a similar way, we will see the universal relation not only the $f$- and $w_1$-modes but also the $p_1$-modes for a wider range of $\Lambda$, adopting various realistic EOSs.
In particular, we will show that the universal relations for the $f$-mode frequency and damping rate derived in this study can predict the corresponding values with less than $1\%$ accuracy.
We note that the resultant universal relations are valid only for cold neutron stars that are described by barotropic EOSs. That is, they may not be valid, for example, for hypermassive neutron stars produced through the binary merger due to the extra degrees of freedom, e.g., temperature, attributed to the EOS.

This manuscript is organized as follows. In Sec. \ref{sec:EOS}, we show the neutron star models considered in this study, where we show the relation between the stellar compactness and $\Lambda$. In Sec. \ref{sec:Oscillation}, we derive the universal relations by showing the frequency and damping rate as a function of $\Lambda$ for various EOSs. Then, in Sec. \ref{sec:Conclusion}, we conclude this study. Unless otherwise mentioned, we adopt geometric units in the following, $c=G=1$, where $c$ denotes the speed of light, and the metric signature is $(-,+,+,+)$.

\section{Neutron star models}
\label{sec:EOS}

To construct the neutron star models, which become the background models for linear analysis, one has to prepare the EOS for neutron star matter. In this study, we adopt the same EOSs as in Ref. \cite{Sotani21}, i.e., the EOSs based on the relativistic framework, DD2~\cite{DD2}, Miyatsu~\cite{Miyatsu}, and Shen~\cite{Shen}; the EOSs with the Skyrm-type interaction, FPS~\cite{FPS}, SKa~\cite{SKa}, SLy4~\cite{SLy4}, and SLy9~\cite{SLy9}; and the EOS constructed with the variational method, Togashi \cite{Togashi17}. We remark that all EOSs adopted here is the unified EOS, i.e., the EOS for the crustal and core region of the neutron star can be constructed with the same framework. The EOS parameters for the EOSs adopted in this study  are listed in Table \ref{tab:EOS} together with the maximum mass of the neutron star constructed with each EOS, where $K_0$ and $L$ are the incompressibility of the symmetric nuclear matter and the density-dependence of the nuclear symmetry energy, and $\eta$ is the combination of $K_0$ and $L$ as $\eta\equiv (K_0L^2)^{1/3}$ \cite{SIOO14}. With the auxiliary parameter $\eta$, one can estimate the properties of the low-mass neutron stars \cite{SIOO14,SSB16} and also discuss the maximum mass \cite{Sotani17,SK17}. 

\begin{table}
\caption{EOS parameters adopted in this study, $K_0$, $L$, and $\eta$. The maximum mass, $M_{\rm max}/M_\odot$, for the neutron star and the dimensionless tidal deformability, $\Lambda_{1.4}$, for the $1.4M_\odot$ neutron star constructed with each EOS are also listed.} 
\label{tab:EOS}
\begin {center}
\begin{tabular}{cccccc}
\hline\hline
EOS & $K_0$ (MeV) & $L$ (MeV) & $\eta$ (MeV) & $M_{\rm max}/M_\odot$ & $\Lambda_{1.4}$  \\
\hline
DD2
 & 243 & 55.0  & 90.2  & 2.41 & 774.8 \\ 
Miyatsu
 & 274 &  77.1 & 118  & 1.95 & 601.0 \\
Shen
\ & 281 & 111  & 151  &  2.17 & 1104.0 \\  
FPS
 & 261 & 34.9 & 68.2  & 1.80 & 182.0 \\  
SKa
 & 263 & 74.6 & 114 & 2.22 & 618.0 \\ 
SLy4
 & 230 & 45.9 &  78.5 & 2.05 & 321.7 \\ 
SLy9
 & 230 & 54.9 &  88.4 & 2.16 & 469.3 \\  
Togashi
 & 245  & 38.7  & 71.6 & 2.21 & 309.2  \\ 
\hline \hline
\end{tabular}
\end {center}
\end{table}

\begin{figure}[tbp]
\begin{center}
\includegraphics[scale=0.6]{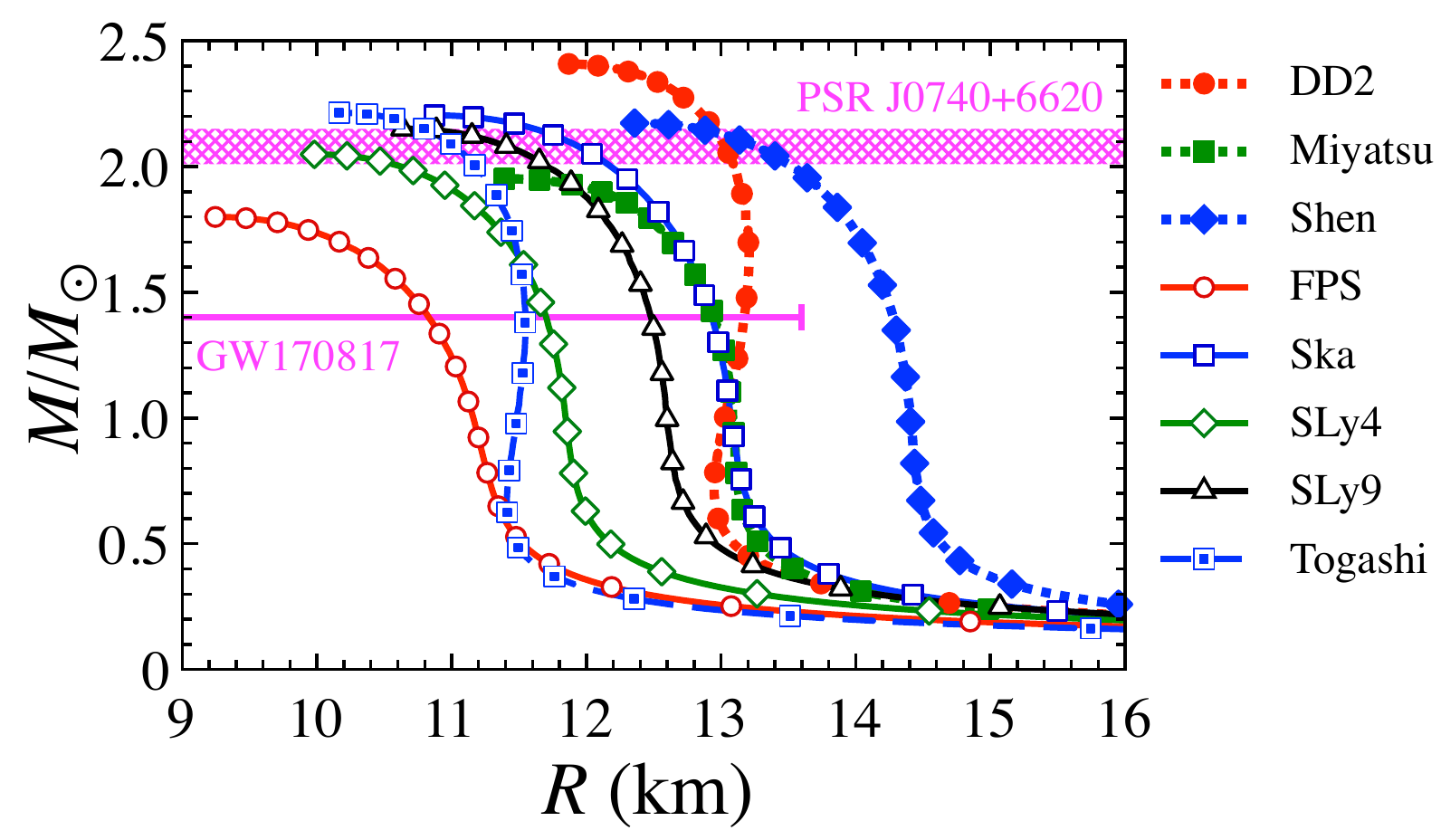} 
\end{center}
\caption{
Neutron star mass and radius relation for various EOSs. For reference, the maximum mass observed so far, i.e., PSR J0740+6620, and the $1.4M_\odot$ neutron star radius constrained from GW170817, i.e., $R_{1.4}\le 13.6$ km, are also shown. 
}
\label{fig:MR}
\end{figure}

The neutron star mass and radius relation constructed with such EOSs are shown in Fig. \ref{fig:MR}, where for reference we also show the maximum mass of neutron star observed so far, 
i.e., $M= 2.08^{+0.07}_{-0.07}M_\odot$
for PSR J0740+6620 \cite{C20,F21}, and the $1.4M_\odot$ neutron star radius constrained from the gravitational wave observation, GW170817, i.e., $R_{1.4} \le 13.6$ km \cite{Annala18}. On the other hand, the terrestrial experiments give us the constraint on $K_0$ and $L$, e.g., $K_0=230\pm 40$ MeV \cite{KM13} and $L\simeq 58.9\pm 16$ MeV \cite{Li19}. Considering these astrophysical and experimental constraints, some of EOSs adopted in this study may be ruled out, but we still adopt them to see the EOS dependence in wide parameter range.

\begin{figure}[tbp]
\begin{center}
\includegraphics[scale=0.6]{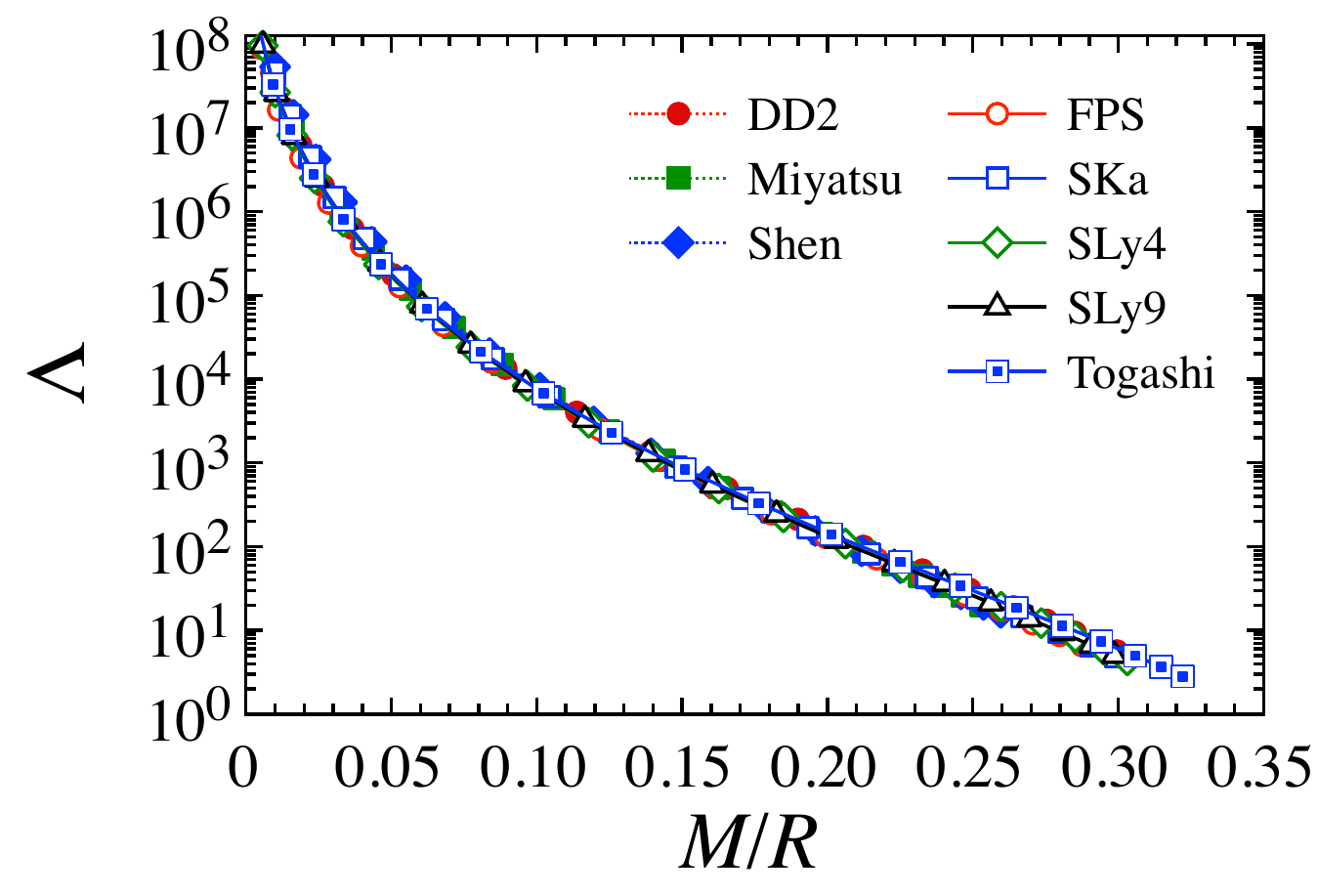} 
\end{center}
\caption{
Relation between the dimensionless tidal deformability $\Lambda$ and compactness $M/R$ for various EOSs. We note that we consider not only the canonical neutron star models but also the neutron star model with quite low compactness such as $M/R\sim 0.01$, which corresponds to $\Lambda\sim 10^8$, in this study.
}
\label{fig:Lam}
\end{figure}

In addition, the dimensionless tidal deformability $\Lambda$ is another important quantity, which could be determined by the gravitational wave observations during the inspiral phase just before the coalescence of binary neutron star. $\Lambda$ is related to the dimensionless quadrupole tidal Love number $k_2$ as
\begin{equation}
 \Lambda = \frac{2}{3}k_2{\cal C}^{-5}, \label{eq:Lam}
\end{equation}
where ${\cal C}$ is the stellar compactness, i.e., ${\cal C}\equiv M/R$, while $k_2$ can be calculated according to Refs. \cite{Hinderer08,Malik18}, as
\begin{eqnarray}
  k_2 &=& \frac{8{\cal C}^5}{5{\cal D}}(1-2{\cal C})^2\left[2 + 2{\cal C}(y_R-1)-y_R\right], \\
  {\cal D} &=& 2{\cal C}\left[6-3y_R + 3{\cal C}(5y_R-8)\right] \nonumber \\
      && +4{\cal C}^3\left[13-11y_R +{\cal C}(3y_R-2)+ 2{\cal C}^2(1+y_R)\right]  \nonumber \\
      && + 3(1-2{\cal C})^2\left[2-y_R + 2{\cal C}(y_R-1)\right]\ln(1-2{\cal C}). \label{eq:k2}
\end{eqnarray}
In this equation, $y_R$ is the surface value of $y$, i.e., $y_R\equiv y(R)$, which can be determined by integrating the following differential equation with respect to $y(r)$ from the center to the surface with the central boundary condition of $y(0)=2$.
\begin{equation}
  r\frac{dy(r)}{dr} = - y(r)^2 - y(r)F(r) - r^2Q(r), \label{eq:dy} 
\end{equation}
with
\begin{gather}
  F(r) = \frac{r-4\pi r^3(\varepsilon -p)}{r-2m}, \\
  Q(r) = \frac{4\pi r}{r-2m}\left(5\varepsilon + 9p + \frac{\varepsilon + p}{\partial p/\partial \varepsilon} 
     - \frac{6}{4\pi r^2}\right)
     - 4\left(\frac{m+4\pi r^3p}{r^2-2mr}\right)^2,
\end{gather}
where $m$, $\varepsilon$, and $p$ denote the mass enclosed within the radius $r$, energy density, and pressure, respectively. In Fig. \ref{fig:Lam} we show the dimensionless tidal deformability $\Lambda$ as a function of the stellar compactness for various EOSs. From this figure, as discussed in Refs. \cite{Yagi14,GGRB21}, one can observe that the dimensionless tidal deformability can be expressed as the stellar compactness almost independently of the EOSs. For reference, the values for the $1.4M_\odot$ neutron star are listed in Table \ref{tab:EOS}.

\section{Quasinormal modes and universal relations}
\label{sec:Oscillation}

On the background models of the static, spherically symmetric neutron star discussed in the previous section, the gravitational wave frequencies are determined by solving the eigenvalue problem, where the frequencies become complex, i.e., the real and imaginary parts correspond to the oscillation frequency and damping rate, respectively. So, this kind of frequencies is sometimes called as quasinormal modes. The concrete perturbation equations and the imposed boundary conditions are the same as in Refs. \cite{STM2001,ST2020}, where how to deal with the boundary condition at the spacial infinity is also discussed. Then, as in the previous study, we focus on only the $\ell=2$ modes in this study, because they are considered to be energetically dominant in the gravitational wave radiation from neutron stars.

\begin{figure*}[tbp]
\begin{center}
\includegraphics[scale=0.6]{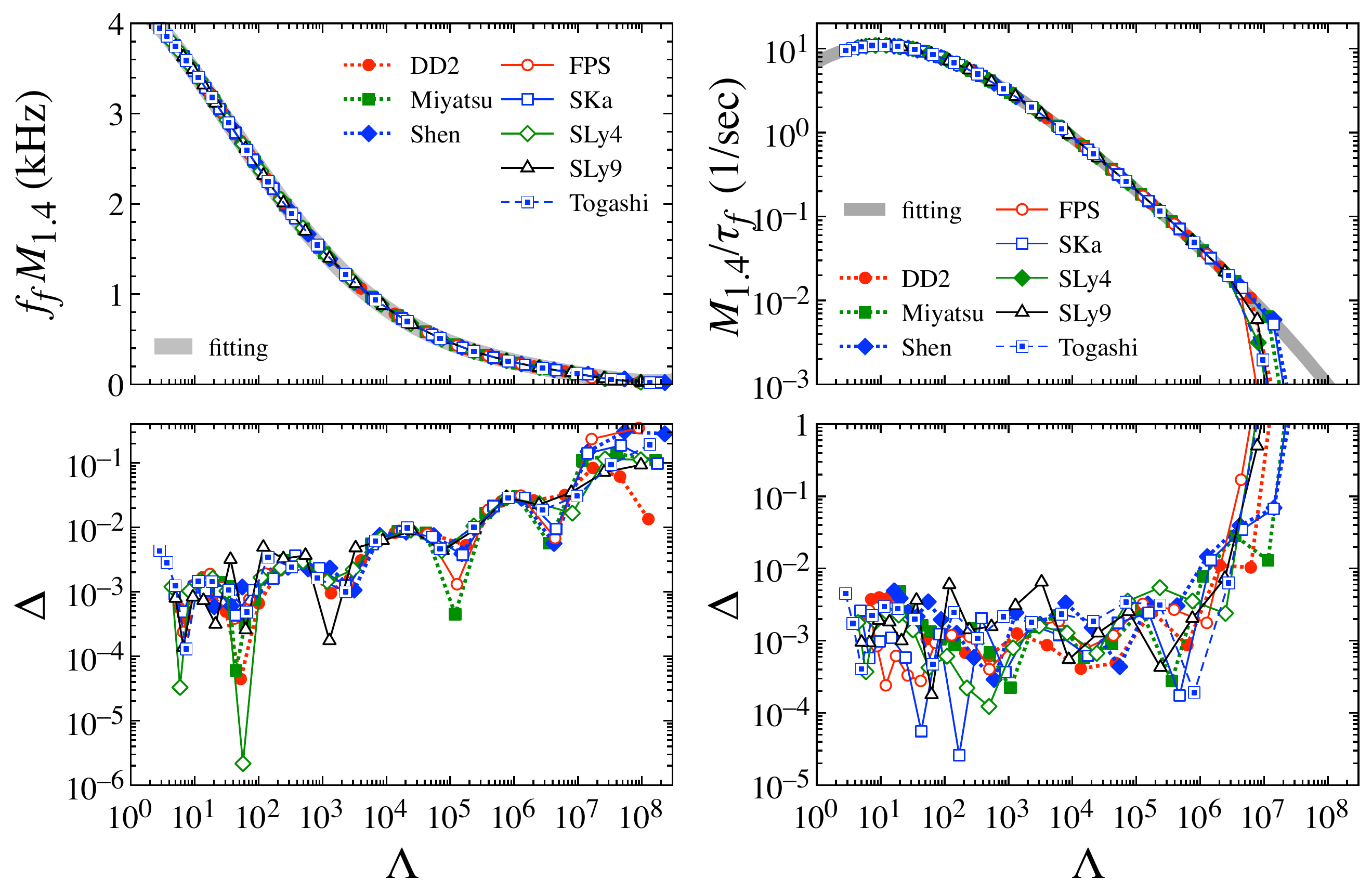} 
\end{center}
\caption{
The $f$-mode frequency multiplied with $M_{1.4}$ (top-left panel) and its damping rate multiplied with $M_{1.4}$ (top-right panel) are shown as a function of $\Lambda$ for various EOSs, where the fitting formulae given by Eqs. (\ref{eq:fit_ff}) -- (\ref{eq:fit_fdamp1}) are also shown with the thick-solid lines. In the bottom panels, the relative deviation between the calculated values and the values predicted with the fitting formulae are shown. 
}
\label{fig:fmode}
\end{figure*}

First, in Fig. \ref{fig:fmode} we show the $f$-mode frequencies, $f_f$, multiplied with the normalized neutron star mass (top-left panel) and their damping rate, $1/\tau_f$, with the damping time $\tau_f$, multiplied with the normalized neutron star mass (top-right panel) are shown as a function of $\Lambda$ for various EOSs. From this figure, one can see that these properties are almost independent of the adopted EOS and can be universally expressed as a function of $\Lambda$. In fact, we can derive the fitting formulae given by 
\begin{gather}
 f_f M_{1.4}\ ({\rm kHz}) = 4.2590 - 0.47874x -0.45353x^2 + 0.14439x^3 - 0.016194x^4 + 0.00064163x^5, \label{eq:fit_ff} \\
 M_{1.4}/\tau_f\ ({\rm 1/sec}) = 10^{g_f(x)}, \label{eq:fit_fdamp} \\ 
 g_f(x) = 0.82691 + 0.45894x - 0.27948x^2 + 0.036480x^3 -0.0025177x^4 + 6.2574\times 10^{-5}x^5, \label{eq:fit_fdamp1}
\end{gather}
where $M_{1.4}\equiv M/(1.4M_\odot)$ and $x=\log_{10}(\Lambda)$, and the predicted values with these universal relations are shown with thick-solid lines in the corresponding panels. The bottom panels show the relative deviation calculated with
\begin{equation}
  \Delta = \frac{|{\cal A} - {\cal A}_{\rm fit}|}{{\cal A}}, \label{eq:devi}
\end{equation}
where $A$ denotes the values of $f_fM_{1.4}$ or $M_{1.4}/\tau_f$ determined by solving the eigenvalue problem, while $A_{\rm fit}$ denotes their values predicted with the fitting formulae. From this figure, one can observe that $f_fM_{1.4}$ and $M_{1.4}/\tau_f$ are estimated with less than $1\%$ accuracy for the canonical neutron star models, whose $\Lambda$ is in the range of $\Lambda \lesssim 10^3$. Considering the universal relations for $f_fM_{1.4}$ and $M_{1.4}/\tau_f$ as a function of $M/R$ derived in Ref. \cite{Sotani21}, which respectively predict a few \% accuracy and $\sim 5\%$ accuracy, the new universal relations derived in this study seem to be more useful. We remark that the reason why the accuracy of the universal relation for low-mass neutron star, e.g., $\Lambda \gsim 10^6$, becomes so bad comes from the avoided crossing between the $f$- and $p_1$-modes, where the behavior of the frequencies and damping rate changes \cite{Sotani21}. For example, as shown in Fig. 8 in Ref. \cite{Sotani21}, the $f$-mode frequency bends at the point where the avoided crossing between the $f$- and $p_1$-modes occurs, while stellar mass does not dramatically change around this point. In addition, we note that we check the other possibility for the universal relation as a function of $\Lambda$, e.g., $f_fR_{10}$, $f_f/u_c$, $R_{10}/\tau_f$, and $1/(u_c\tau_f)$, where $R_{10}$ and $u_c$ are defined by $R_{10}\equiv R/(10\ {\rm km})$ and $u_c\equiv \rho_c/\rho_0$ with the central density $\rho_c$ and the nuclear saturation density $\rho_0$. But, we find that the universal relation given by Eqs. (\ref{eq:fit_ff}) and (\ref{eq:fit_fdamp}) are more accurate than these other possible combinations.

\begin{figure*}[tbp]
\begin{center}
\includegraphics[scale=0.6]{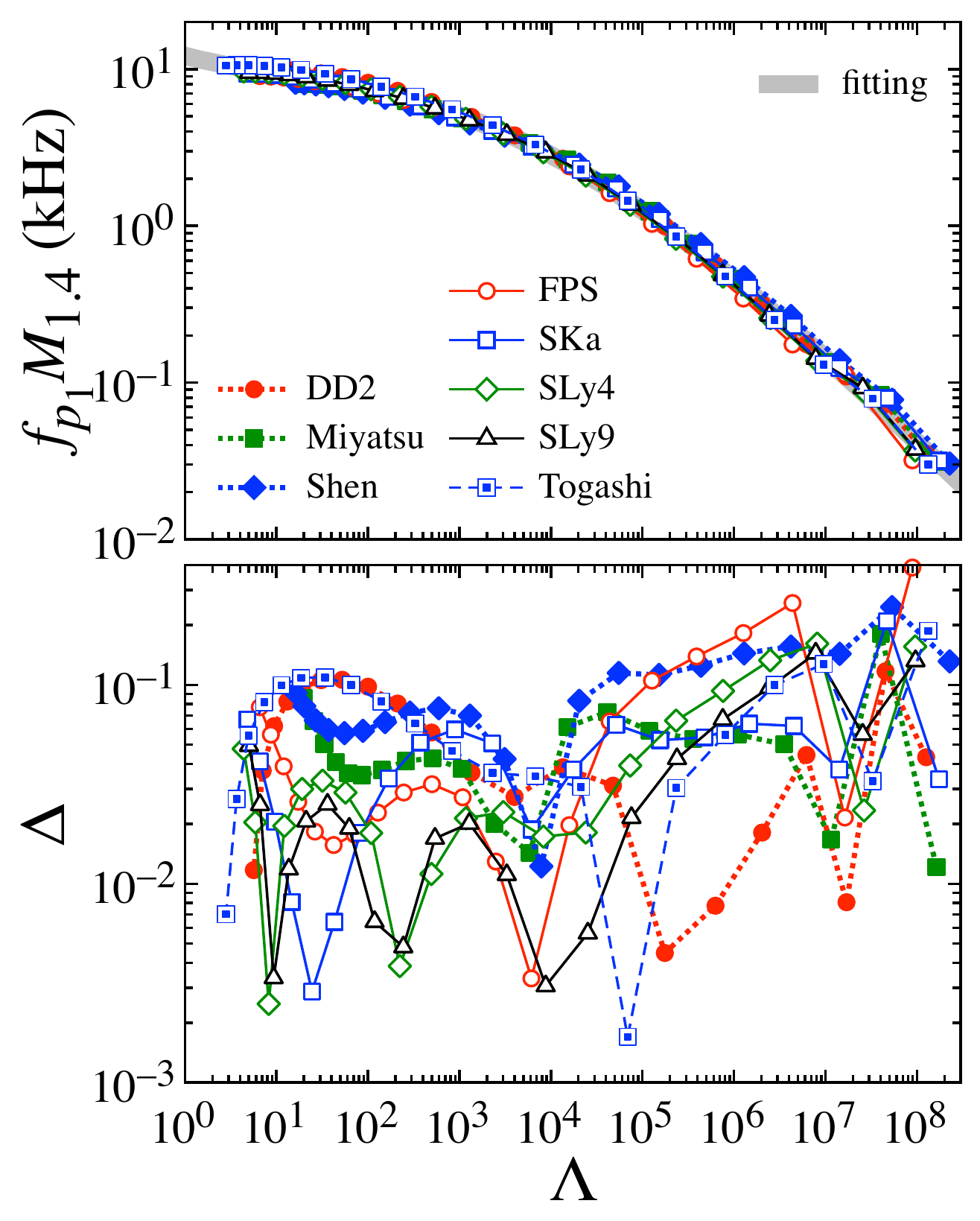} 
\end{center}
\caption{
The $p_1$-mode frequency multiplied with $M_{1.4}$ (top-left panel) is shown as a function of $\Lambda$ for various EOSs, where the fitting formulae given by Eqs. (\ref{eq:fit_fp1}) and (\ref{eq:fit_fp11}) is also shown with the thick-solid lines. In the bottom panel, the relative deviation between the calculated values and the values predicted with the fitting formula is shown. 
}
\label{fig:p1mode}
\end{figure*}

Next, in Fig. \ref{fig:p1mode} we show the $p_1$-mode frequencies, $f_{p_1}$, multiplied with the normalized neutron star mass is shown as a function of $\Lambda$ in the top panel. Again, one can see that these quantities are almost independent of the adopted EOSs and universally expressed as a function of $\Lambda$, such as
\begin{gather}
 f_{p1} M_{1.4}\ ({\rm kHz}) = 10^{g_{p_1}(x)}, \label{eq:fit_fp1} \\
 g_{p_1}(x) = 1.0853 - 0.15527x + 0.062266x^2 - 0.023666x^3 + 0.0022713x^4 - 7.3071\times 10^{-5}x^5, \label{eq:fit_fp11}
\end{gather}
with which the expected values are also shown with thick-solid line. In the bottom panel, we show the relative deviation of the calculated frequencies from the fitting formula in the similar way to the case of the $f$-modes. With this universal relation, one can predict $f_{p_1}M_{1.4}$ within $\sim 10\%$ accuracy for canonical neutron star, which is more or less similar accuracy with the universal relation as a function of $M/R$ derived in Ref. \cite{Sotani21}. On the other hand, as shown in Ref. \cite{AK1998,Sotani21} it is difficult to derive a kind of universal relation for the damping rate of the $p_1$-modes, which strongly depends on the EOSs.

\begin{figure*}[tbp]
\begin{center}
\includegraphics[scale=0.6]{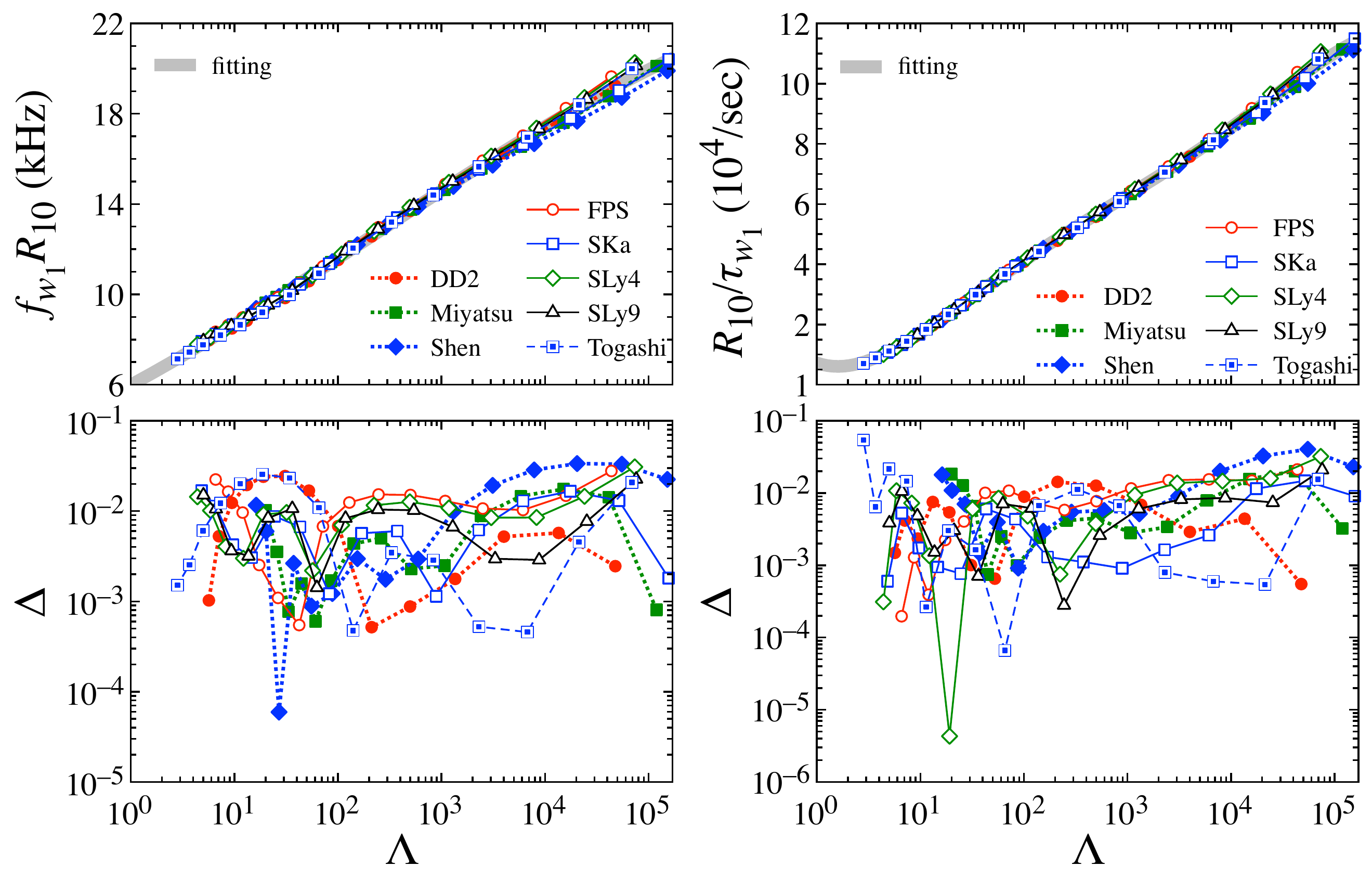} 
\end{center}
\caption{
The $w_1$-mode frequency multiplied with $R_{10}$ (top-left panel) and its damping rate multiplied with $R_{10}$ (top-right panel) are shown as a function of $\Lambda$ for various EOSs, where the fitting formulae given by Eqs. (\ref{eq:fit_fw1}) and (\ref{eq:fit_w1damp}) are also shown with the thick-solid lines. In the bottom panels, the relative deviation between the calculated values and the values predicted with the fitting formulae are shown. 
}
\label{fig:w1mode}
\end{figure*}

Furthermore, in Fig. \ref{fig:w1mode} we show the $w_1$-mode frequencies, $f_{w_1}$, multiplied with the normalized neutron star radius (top-left panel) and their damping rate, $1/\tau_{w_1}$, with the damping time $\tau_{w_1}$ multiplied with the normalized neutron star radius (top-right panel) are shown as a function of $\Lambda$ for various EOSs.
We note that, unlike the $f$- and $p_1$-modes, we can numerically determine the $w_1$-mode quasinormal modes only for the neutron star models with $M/R\gsim 0.05$, which corresponds to $\Lambda \lsim 10^5$.  
In the same fashion as the $f$- and $p_1$-modes, one can observe that these quantities are almost independent of the adopted EOSs and we can derive the universal relations as
\begin{gather}
  f_{w_1}R_{10}\ ({\rm kHz}) = 5.9881 + 2.4830x + 0.22713x^2 - 0.033050x^3,   \label{eq:fit_fw1} \\ 
  R_{10}/\tau_{w_1}\ ({\rm 1/sec}) = (0.72183 -1.1866x + 3.3208x^2 -1.3880x^3 +0.25982x^4 -0.017818x^5)\times 10^4, 
      \label{eq:fit_w1damp}
\end{gather}
where $R_{10}\equiv R/(10\ {\rm km})$. In the bottom panes, the corresponding relative deviation of the calculated values from the values predicted from the universal relations are shown. From this figure, the universal relations as a function of $\Lambda$ can reasonably predict the values of $f_{w_1}R_{10}$ and $R_{10}/\tau_{w_1}$, but their accuracy is more or less similar to that with the universal relation as a function of $M/R$ derived in Ref. \cite{Sotani21}. Anyway, the detection of $w_1$-mode gravitational waves must be quite more difficult, due to their high damping rate.

\section{Conclusion}
\label{sec:Conclusion}

In this study, focusing on the feature that the relation between the neutron star compactness and the dimensionless tidal deformability hardly depends on the EOSs, we derive the universal relations predicting the $f$- and $p_1$-mode frequencies and the $f$-mode damping rate multiplied with the normalized neutron star mass, and the $w_1$-mode frequency and its damping rate multiplied with the normalized neutron star radius as a function of the dimensionless tidal deformability. With the universal relations for the $f$-mode, one can predict the properties more accurate than that with the universal relation as a function of compactness, while the universal relations for the $p_1$- and $w_1$-modes are more or less similar predictability to the previous ones. With these universal relations derived in this study, one may discuss the gravitational waves from the neutron stars in the binary neutron star system, where one has a chance to know the value of the dimensionless tidal deformability. In this study, we adopt the EOSs, whose parameters are in relatively wide range, but if one adopts only the EOSs constrained from the astronomical observations and terrestrial experiments in more narrow parameter range and modifies the coefficients in the universal relations, their predictability may become more accurate. Finally, we remark that the present work can be helpful for detecting gravitational waves from quasi-normal oscillation modes of neutron stars, and it will also help improve the tidal deformability, especially its lower limit.

\acknowledgments

This work is supported in part by Japan Society for the Promotion of Science (JSPS) KAKENHI Grant Numbers 
JP19KK0354,  
JP20H04753,  and 
JP21H01088,  
and by Pioneering Program of RIKEN for Evolution of Matter in the Universe (r-EMU).



\begin{thebibliography}{999}
\bibitem{gw170817} 
   B. P. Abbott et al. (The LIGO Scientific Collaboration and the Virgo Collaboration), Phys. Rev. Lett. {\bf 119}, 161101 (2017).

\bibitem{GW190425} 
   B. P. Abbott et al. (The LIGO Scientific Collaboration and the Virgo Collaboration), Astrophys. J. {\bf 892}, L3 (2020).

\bibitem{gw170817a} 
   B. P. Abbott et al. (The LIGO Scientific Collaboration and the Virgo Collaboration), Phys. Rev. Lett. {\bf 121}, 161101 (2018).

\bibitem{ST83}
   S. L. Shapiro and S. A. Teukolsky, {\it Black Holes, White Dwarfs, and Neutron Stars: The Physics of Compact Objects}  (Wiley-Interscience, New York, 1983).

\bibitem{D10} 
   P. Demorest, T. Pennucci, S. Ransom, M. Roberts, and J. Hessels, Nature {\bf 467}, 1081 (2010).

\bibitem{A13} 
   J. Antoniadis {\it et al.}, Science {\bf 340}, 6131 (2013).

\bibitem{C20}    
   H. T. Cromartie {\it et al.}, Nature Astronomy {\bf 4}, 72 (2020).

\bibitem{Annala18}  
   E. Annala, T. Gorda, A. Kurkela, and A. Vuorinen, Phys. Rev. Lett. {\bf 120}, 172703 (2018).

\bibitem{PFC83} 
   K. R. Pechenick, C. Ftaclas, and J. M. Cohen, Astrophys. J. {\bf 274}, 846 (1983).

\bibitem{LL95} 
   D. A. Leahy and L. Li, Mon. Not. R. Astron. Soc. {\bf 277}, 1177 (1995).

\bibitem{PG03} 
   J. Poutanen and M. Gierlinski, Mon. Not. R. Astron. Soc. {\bf 343}, 1301 (2003).

\bibitem{PO14} 
   D. Psaltis and F. \"{O}zel, Astrophys. J. {\bf 792}, 87 (2014). 

\bibitem{SM18} 
   H. Sotani and U. Miyamoto, Phys. Rev. D {\bf 98}, 044017 (2018); {\bf 98}, 103019 (2018).

\bibitem{Sotani20a} 
   H. Sotani, Phys. Rev. D {\bf 101}, 063013 (2020).

\bibitem{Riley19} 
   T. E. Riley {\it et al.}, Astrophys. J.  {\bf 887}, L21 (2019).
   
\bibitem{Miller19} 
   M. C. Miller {\it et al.}, Astrophys. J.  {\bf 887}, L24 (2019).

\bibitem{Riley21} 
   T. E. Riley {\it et al.}, Astrophys. J.  {\bf 918}, L27 (2021).
   
\bibitem{Miller21} 
   M. C. Miller {\it et al.}, Astrophys. J.  {\bf 918}, L28 (2021).

\bibitem{GNHL2011}
   M. Gearheart, W. G. Newton, J. Hooker, and B. -A. Li, Mon. Not. R. Astron. Soc. {\bf 418}, 2343 (2011).
   
\bibitem{SNIO2012}
   H. Sotani, K. Nakazato, K. Iida, and K. Oyamatsu, Phys. Rev. Lett. {\bf 108}, 201101 (2012);
   Mon. Not. R. Astron. Soc. {\bf 428}, L21 (2013); {\bf 434}, 2060 (2013).

\bibitem{SIO2016}
   H. Sotani, K. Iida, and K. Oyamatsu, New Astron. {\bf 43}, 80 (2016);
   Mon. Not. R. Astron. Soc. {\bf 464}, 3101 (2017); {\bf 479}, 4735 (2018); 
   {\bf 489}, 3022 (2019).

 
\bibitem{AK1996}
   N. Andersson and K. D. Kokkotas, Phys.\ Rev.\ Lett.\ {\bf 77}, 4134 (1996).

\bibitem{AK1998}
   N. Andersson and K. D. Kokkotas, Mon.\ Not.\ R. Astron.\ Soc.\ {\bf 299}, 1059 (1998).

\bibitem{Benhar99}
   O. Benhar, E. Berti, and V. Ferrari, Mon.\ Not.\ R. Astron.\ Soc.\ {\bf 310}, 797 (1999).

\bibitem{STM2001}
   H. Sotani, K. Tominaga, and K. I. Maeda, Phys.\ Rev.\ D {\bf 65}, 024010 (2001).

\bibitem{SH2003}
   H. Sotani and T. Harada, Phys.\ Rev.\ D {\bf 68}, 024019 (2003);
   H. Sotani, K. Kohri, and T. Harada, {\it ibid}.\ {\bf 69}, 084008 (2004).

\bibitem{Benhar04}
   O. Benhar, V. Ferrari and L. Gualtieri, Phys.\ Rev.\ D {\bf 70}, 124015 (2004).

\bibitem{TL2005}
   L. K. Tsui and P. T. Leung, Mon.\ Not.\ R. Astron.\ Soc.\ {\bf 357}, 1029 (2005).

\bibitem{Lau10}
   H. K. Lau, P. T. Leung, and L. M.  Lin, Astrophys. J. {\bf 714}, 1234 (2010).

\bibitem{SYMT2011}
   H. Sotani, N. Yasutake, T. Maruyama, and T. Tatsumi, Phys.\ Rev.\ D {\bf 83} 024014 (2011).

\bibitem{PA2012}
   A. Passamonti and N. Andersson, Mon.\ Not.\ R. Astron.\ Soc.\ {\bf 419}, 638 (2012).

\bibitem{BGN13}
   J. L. Bl\'{a}zquez-Salcedo, L. M. Gonz\'{a}lez-Romero, and F. Navarro-L\'{e}rida, Phys.\ Rev.\ D {\bf 87}, 104042 (2013).

\bibitem{DGKK2013}
   D. D. Doneva, E. Gaertig, K. D. Kokkotas, and C. Kr\"{u}ger, Phys.\ Rev.\ D {\bf 88}, 044052 (2013).

\bibitem{CdK15}
   C. Chirenti, G. H. de Souza, and W. Kastaun, Phys.\ Rev.\ D {\bf 91}, 044034 (2015).

\bibitem{Sotani20b}
   H. Sotani, Phys. Rev. D {\bf 102}, 063023 (2020); 103021 (2020).

\bibitem{Sotani21}
   H. Sotani, Phys. Rev. D {\bf 103}, 123015 (2021).

\bibitem{FMP2003}
   V. Ferrari, G. Miniutti, and J. A. Pons, Mon. Not. R. Astron. Soc. {\bf 342}, 629 (2003).

\bibitem{FKAO2015}
   J. Fuller, H. Klion, E. Abdikamalov, and C. D. Ott, Mon.\ Not.\ R. Astron.\ Soc.\ {\bf 450}, 414 (2015).

\bibitem{ST2016}
   H. Sotani and T. Takiwaki, Phys.\ Rev.\ D {\bf 94}, 044043 (2016); {\bf 102}, 023028 (2020); 
   Mon. Not. R. Astron. Soc. {\bf 498}, 3503 (2020).

\bibitem{SKTK2017}
   H. Sotani, T. Kuroda, T. Takiwaki, and K. Kotake, Phys.\ Rev.\ D {\bf 96}, 063005 (2017).

\bibitem{MRBV2018}
  V. Morozova, D. Radice, A. Burrows, and D. Vartanyan, Astrophys. J. {\bf 861}, 10 (2018).

\bibitem{SKTK2019}
   H. Sotani, T. Kuroda, T. Takiwaki, and K. Kotake, Phys.\ Rev.\ D {\bf 99}, 123024 (2019).

\bibitem{TCPOF19}
   A. Torres-Forn\'{e}, P. Cerd\'{a}-Dur\'{a}n, A. Passamonti, M. Obergaulinger, and J. A. Font, Mon. Not. R. Astron. Soc. {\bf 482}, 3967 (2019).

\bibitem{SS2019}
   H. Sotani and K. Sumiyoshi, Phys.\ Rev.\ D {\bf 100}, 083008 (2019); Mon. Not. R. Astron. Soc. {\bf 507}, 2766 (2021).

\bibitem{ST2020}
   H. Sotani and T. Takiwaki, Phys.\ Rev.\ D {\bf 102}, 063025 (2020).

\bibitem{STT2021}
   H. Sotani, T. Takiwaki, and H. Togashi, preprint (arXiv:2110.03131).

\bibitem{Chan14}
   T. K. Chan, Y.-H. Sham, P. T. Leung, and L.-M. Lin,  Phys. Rev. D {\bf 90}, 124023 (2014).

\bibitem{Wen19}
   D. H. Wen, B. A. Li, H. Y. Chen, and N. B. Zhang, Phys.\ Rev.\ C {\bf 99}, 045806 (2019).

\bibitem{MG19}
   J. Mena-Fern\'{a}ndez and L. M. Gonz\'{a}lez-Romero, Phys.\ Rev.\ C {\bf 99}, 104005 (2019).

\bibitem{Benitex21}
   E. Benitez, J. Weller, V. Guedes, C. Chirenti, and M. C. Miller, Phys. Rev. D {\bf 103}, 023007 (2021).
   

\bibitem{LBN21}  
   G. Lioutas, A. Bauswein, and N. Stergioulas, Phys. Rev. D {\bf 104}, 043011 (2021).





   

   
\bibitem{DD2}  
   S. Typel, Phys. Rev. C {\bf 89}, 064321 (2014).

\bibitem{Miyatsu}  
   T. Miyatsu, S. Yamamuro, and K. Nakazato, Astrophys. J. {\bf 777}, 4 (2013).

\bibitem{Shen}  
   H. Shen, H. Toki, K. Oyamatsu, and K. Sumiyoshi, Nucl. Phys. {\bf A637}, 435 (1998).

\bibitem{FPS}  
   C. P. Lorenz, D. G. Ravenhall, and C. J. Pethick, Phys. Rev. Lett. {\bf 70}, 379 (1993).

\bibitem{SKa}  
   H. S. K\"{o}hler, Nucl. Phys. A {\bf 258}, 301 (1976).
   
\bibitem{SLy4}  
   F. Douchin and P. Haensel, Astron. Astrophys. {\bf 380}, 151 (2001).

\bibitem{SLy9}  
   E. Chabanat, {\it Interactions effectives pour des conditions extremes d'isospin}, Ph.D. thesis, University Claude Bernard Lyon-I (1995).

\bibitem{Togashi17}  
   H. Togashi, K. Nakazato, Y. Takehara, S. Yamamuro, H. Suzuki, and M. Takano, Nucl. Phys. A {\bf 961}, 78 (2017).
  

\bibitem{SIOO14} 
   H. Sotani, K. Iida, K. Oyamatsu, and A. Ohnishi, Prog. Theor. Exp. Phys. {\bf 2014}, 051E01 (2014).
  
\bibitem{SSB16} 
   H. O. Silva, H. Sotani, and E. Berti, Mon. Not. R. Astron. Soc. {\bf 459}, 4378 (2016).   

\bibitem{Sotani17}  
   H. Sotani, Phys. Rev. C {\bf 95}, 025802 (2017).

\bibitem{SK17}  
   H. Sotani and K. D. Kokkotas, Phys. Rev. D {\bf 95}, 044032 (2017).
   

\bibitem{F21}
   E. Fonseca et al., Astrophys. J. {\bf 915}, L12 (2021).


\bibitem{KM13}
   E. Khan and J. Margueron, Phys. Rev. C {\bf 88}, 034319 (2013).

\bibitem{Li19}
   B. A. Li, P. G. Krastev, D. H. Wen, and N. B. Zhang, Euro. Phys. J. A {\bf 55}, 117 (2019).

\bibitem{Hinderer08}
  T. Hinderer, The Astrophysical Journal 677, 1216 (2008).

\bibitem{Malik18}
  T. Malik, N. Alam, M. Fortin, C. Provid\^{e}ncia, B. K. Agrawal, T. K. Jha, Bharat Kumar, S. K. Patra, Phys. Rev. C 98, 035804 (2018)

\bibitem{Yagi14}
   K. Yagi, Phys. Rev. D {\bf 89}, 043011 (2014).

\bibitem{GGRB21}
   D. A. Godzieba, R. Gamba, D. Radice, and S. Bernuzzi, Phys. Rev. D {\bf 103}, 063036 (2021).


\end{thebibliography}

\end{document}